\documentclass{elsart4-1}

\usepackage{amssymb}

\usepackage[english,french]{babel}

\newtheorem{e-proposition}[theorem]{Proposition}

\newtheorem{e-definition}[theorem]{Definition\rm}

\renewcommand{\theequation}{\arabic{equation}}
\def\og{\leavevmode\raise.3ex\hbox{$\scriptscriptstyle\langle\!\langle$~}}
\def\fg{\leavevmode\raise.3ex\hbox{~$\!\scriptscriptstyle\,\rangle\!\rangle$}}

\begin{document}

\begin{frontmatter}


\selectlanguage{english}
\title{Exact scaling transform for a unitary quantum gas in a time
dependent harmonic potential}

\vspace{-2.6cm}

\selectlanguage{french}
\title{ \'Evolution d'un gaz quantique unitaire dans un potentiel
harmonique variable : solution par changement d'\'echelle }


\selectlanguage{english}
\author{Yvan Castin}
\ead{yvan.castin@lkb.ens.fr}

\address{Laboratoire Kastler Brossel, 
\'Ecole normale sup\'erieure, 24 rue Lhomond, 75 231 Paris, France}

\begin{abstract}
A unitary quantum gas is a gas of quantum particles with a binary interaction
of infinite scattering length and negligible 
range. It has been produced in recent experiments with gases of fermionic
atoms by means of a Feshbach resonance. 
Using the Fermi pseudo-potential model for the atomic interaction,
we show that the time evolution of such a gas in an isotropic three-dimensional
time dependent harmonic trap is exactly given by a gauge
and scaling transform.

\vskip 0.5\baselineskip

\selectlanguage{french}
\noindent{\bf R\'esum\'e}

\vskip 0.5\baselineskip

\noindent
Nous entendons par ``gaz quantique unitaire" une assembl\'ee de particules
dont le mouvement est d\'ecrit quantiquement et qui interagissent par
un potentiel de longueur de diffusion infinie et de port\'ee n\'egligeable
devant leur distance moyenne et leur longueur d'onde thermique. 
Un tel gaz a \'et\'e  produit r\'ecemment
\`a l'aide d'une r\'esonance de Feshbach dans un gaz d'atomes fermioniques.
En mod\'elisant les interactions entre particules par le pseudo-potentiel
de Fermi, nous montrons que l'\'evolution d'un gaz unitaire dans un potentiel
de pi\'egeage harmonique isotrope tridimensionnel de d\'ependence
temporelle quelconque est d\'ecrite
exactement par la composition d'un changement d'\'echelle et
d'une transformation de jauge.

\vskip 0.5\baselineskip
\noindent{\small{\it Mots-cl\'es~:} 
gaz quantique unitaire~; changement d'\'echelle~; potentiel harmonique}
}
\end{abstract}
\end{frontmatter}


\selectlanguage{english}
\section{Introduction}
\label{sec:intro}

Experiments with quantum gases of spin $1/2$ fermionic atoms are currently
making rapid progresses. One of the most
fascinating properties of these fermionic gases is the possibility
to freely tune the sign and the strength 
of the atomic interactions without reducing the lifetime of the sample: 
the value of the $s$-wave scattering length $a$
of two particles with opposite spin components can virtually be adjusted from 
$-\infty$ to $+\infty$ with a Feshbach resonance technique \cite{Feshbach}, 
without inducing any instability of the gas even in
the unitary limit $a=\pm\infty$ \cite{Thomas}.
These systems are still gases, in the sense that the effective range
of the interaction potential is negligible as compared to
the mean interparticle separation and to the thermal de Broglie wavelength. 
We shall take advantage
of this crucial property and model the true interaction potential
by the so-called Fermi 
pseudo-potential \cite{Houches}.

Such stability of the strongly interacting Fermi gases opens up 
fascinating possibilities, 
e.g. the study of the crossover 
between a Bose-Einstein
condensate of dimers (already observed, see \cite{Jin,Grimm,Ketterle,Salomon})
and a BCS condensate of pairs, by passing through
the strongly interacting regime $k_F |a|\gg 1$, where
$k_F$ is the Fermi momentum
\cite{Nozieres,Randeria}. In the
unitary limit $k_F |a|\rightarrow +\infty$, 
the thermodynamic properties of the spatially homogeneous gas 
are universal: they depend only on the Fermi energy and on the temperature.
At zero temperature, the chemical potential of the homogeneous gas
is then $\mu = \eta \mu_0$
where $\eta$ is a pure number and $\mu_0$ is the chemical potential
of the ideal Fermi gas \cite{Thomas,Pandharipande}. 
An accurate measurement of 
$\eta$ would provide a crucial test of many-body
theories \cite{Thomas,Grimm,Salomon}. 

The standard imaging technique used with quantum
gases is to switch off 
the trapping potential, to let the gas expand and to perform a
light absorption imaging of the atomic cloud. 
Such a ballistic expansion acts as a magnifying lens: it was used
to reveal the vortex lattice in a rotating Bose-Einstein condensate
\cite{Dalibard}, and very recently to obtain the value of the universal number $\eta$
for the unitary Fermi gas \cite{Salomon}.
Clearly the interpretation of the time of flight images strongly
relies on a theoretical understanding of the time evolution of
the gas in a time dependent harmonic potential.
In the case of a pure Bose-Einstein condensate in the regime $k_F |a|
\ll 1$, this was achieved starting from the Gross-Pitaevskii equation
by a gauge plus scaling transform \cite{Dum,Shlyapnikov}.
When the Bose or Fermi gas enters the strongly interacting regime
$k_F |a| > 1$, no solution starting from first principles
is available \cite{espoir} and one relies on the hydrodynamic approximation \cite{Stringari}.

Here we consider the idealized case of an isotropic
and harmonic three-dimensional trapping potential.
In the limit of an infinite
scattering length, we show that 
the Fermi pseudo-potential has a scaling invariance
that rigorously allows the use of a gauge plus scaling transform 
similar to the one of \cite{Pitaevskii} to describe
the time evolution of the gas due to an arbitrary
variation of the trapping frequency.

\section{The model based on the Fermi pseudo-potential}
\label{sec:model}
Consider an assembly of $N$ non relativistic particles, with an arbitrary spin.
These particles may be indistinguishable bosons or fermions, or even
be distinguishable. All the particles have the same mass $m$ and interact
{\sl via} the same binary interaction potential independent of
the spin degrees of freedom.
The interaction potential is the Fermi pseudo-potential with
coupling constant $g$ related to the $s$-wave scattering length
$a$ by $g = 4\pi \hbar^2 a/m$.
At this stage, $0< |a| < +\infty$, we shall take the unitary limit
$|a|\rightarrow +\infty$ later.

Let $\psi(\mathbf{r_1},\ldots,\mathbf{r_N})$
be the wavefunction of the gas corresponding to a given (but arbitrary)
spin configuration. The wavefunction $\psi$ then evolves
according to the Schr\"odinger equation:
\begin{equation}
i\hbar \partial_t \psi =
\sum_{i=1}^{N} \left[
-\frac{\hbar^2}{2m} \Delta_{\mathbf{r_i} }
+U(\mathbf{r_i})\right]\psi + 
\sum_{1\leq i < j \leq N} g \delta(\mathbf{r_i}
-\mathbf{r_j}) \psi^{\mathrm{reg} }_{ij}.
\label{eq:schrodinger}
\end{equation}
Here $\Delta_{\mathbf{r_i} }$ is the three-dimensional Laplacian
with respect to the spatial coordinates $\mathbf{r_i}$ 
of particle number $i$, $U$ is the
trapping potential seen by each particle and each
$\psi^{\mathrm{reg} }_{ij}$, the so-called regular
part of $\psi$ in $\mathbf{r_i}=\mathbf{r_j}$,
is the following function of $N-1$ vectors of coordinates:
\begin{equation}
\psi^{\mathrm{reg} }_{ij}(\{\mathbf{r_k},k\neq i,j\},
\mathbf{R_{ij}}) =
\lim_{r_{ij}\rightarrow 0} \frac{\partial}{\partial r_{ij}}\left(r_{ij}
\psi\right)
\end{equation}
where $r_{ij}$ is the norm of 
$\mathbf{r_{ij}}\equiv \mathbf{r_i}-\mathbf{r_j}$ and
where the limit and the partial derivative are taken for fixed positions
$\mathbf{r_k}$ of the $N-2$ particles other than
particles $i$ and $j$, and for a fixed position of the center of mass
of the particles $i$ and $j$,
$\mathbf{R_{ij} } \equiv (\mathbf{r_i} + \mathbf{r_j})/2$.

The domain of the Hamiltonian corresponding to 
the Fermi pseudo-potential is therefore 
not the Hilbert space of the non-interacting gas, but a functional
space with specific boundary conditions for the wavefunction $\psi$.
More precisely, as we now see,
the model amounts to replacing the true interaction potential
by {\sl contact} conditions, i.e. by boundary conditions on $\psi$ when the distance 
$r_{ij}$ between two particles tends
to zero, the wavefunction
$\psi$ otherwise evolving with the interaction free Schr\"odinger equation
\cite{Petrov}. 
As the wavefunction $\psi$ does not contain any delta
singularity, each delta singularity in the interaction term of
Eq.(\ref{eq:schrodinger}) has indeed to be compensated by a delta
singularity in the kinetic energy term. In 3D this implies
that $\psi$ can diverge as $1/r_{ij}$ when $r_{ij}\rightarrow 0$,
a divergence which is still square integrable. 
Two cases can occur:
\begin{itemize}
\item case i): $\lim_{r_{ij}\rightarrow 0} \psi =0$: 
no delta singularity occurs from the kinetic energy operator, and
there is no delta singularity from the interaction term 
as $\psi_{ij}^{\mathrm {reg}}$ vanishes.
For instance, this is the case when particles $i$ and $j$ are indistinguishable 
fermions in the same spin state.
\item case ii): $\psi$ has a $1/r_{ij}$ singularity:
\begin{equation}
\psi = \frac{A}{r_{ij} } + B + O(r_{ij})
\end{equation}
where $A$ and $B$ are still functions of the $\mathbf{r_k}$'s with
$k\neq i,j$, and of
$\mathbf{R_{ij} }$. The regular part of $\psi$ is then
$\psi_{ij}^{\mathrm {reg}} = B$.
Writing the kinetic energy operator for the pair of particles $i,j$
as $\Delta_{\mathbf{r_i} } + \Delta_{\mathbf{r_j} } =
\frac{1}{2} \Delta_{\mathbf{R_{ij}}}+ 2 \Delta_{\mathbf{r_{ij}}}$
and using $\Delta(1/r) = -4\pi \delta(\mathbf{r})$,
we find that the total coefficient of $\delta(\mathbf{r_i}-\mathbf{r_j})$ 
in the right hand side of Eq.(\ref{eq:schrodinger}) vanishes provided
that $ A + a B =0$.
\end{itemize}
A way of summarizing the two cases is then simply to impose the boundary
conditions:
\begin{equation}
\psi(\mathbf{r_1},\ldots,\mathbf{r_N})
= A(\{\mathbf{r_k},k\neq i,j\},\mathbf{R_{ij}})
[r_{ij}^{-1} - a ^{-1}]
+O(r_{ij})
\label{eq:cl}
\end{equation}
the first case corresponding to $A=0$ and the second one to $A\neq 0$.
Having ensured that all the delta singularities cancel in the Schr\"odinger
equation, we can now restrict it to the manifold 
where the positions of the particles
are two by two distinct:
\begin{equation}
i\hbar \partial_t \psi =
\sum_{i=1}^{N} \left[ -\frac{\hbar^2}{2m} \Delta_{\mathbf{r_i}}
+ U(\mathbf{r_i})\right]\psi 
\ \ \ \ \mbox{for} \ \ \ r_{ij} \neq 0, \forall i\neq j.
\label{eq:overD}
\end{equation}
Eq.(\ref{eq:cl}) and Eq.(\ref{eq:overD}) constitute the basis
of our model.

\section{Scaling transform in the unitary limit}
\label{sec:scaling}

We specialize the previous section to the case of the unitary quantum
gases: the scattering length is now infinite, so that we set 
to zero the $1/a$ term
in the boundary conditions (\ref{eq:cl}), to obtain
\begin{equation}
\psi(\mathbf{r_1},\ldots,\mathbf{r_N})
= \frac{A(\{\mathbf{r_k},k\neq i,j\},\mathbf{R_{ij}}) }
{r_{ij} } +O(r_{ij}).
\label{eq:clu}
\end{equation}
Note that  the unitary limit does not look in any way singular 
in our formulation of the model,
as the scattering length only appears through its inverse 
in Eq.(\ref{eq:cl}).
Also we restrict to the case of an isotropic harmonic potential:
\begin{equation}
U(\mathbf{r}) = \frac{1}{2} m \omega^2(t) r^2
\end{equation}
where the oscillation frequency $\omega(t)$ of a particle in
a trap is constant and equal to $\omega_0$ for $t<0$
and has an arbitrary time dependence for $t\geq 0$. 
The ballistic expansion
of the gas mentioned in the introduction corresponds to setting $\omega$
to zero for $t\geq 0$.

We assume that the state vector of the gas for $t<0$ is a steady
state of Schr\"odinger's equation with the energy $E$. The corresponding
wavefunction for the considered spin configuration is $\psi_0$, which
in particular obeys the boundary condition Eq.(\ref{eq:clu}).
Our ansatz for the time-dependent wavefunction, inspired
from \cite{Dum,Shlyapnikov,Pitaevskii}, is then
\begin{equation}
\psi(\mathbf{r_1},\ldots,\mathbf{r_N},t)
= \mathcal{N}(t)
e^{i\sum_{j=1}^N m r_j^2 \dot\lambda(t)/2\hbar \lambda(t) }
\psi_0(\mathbf{r_1}/ \lambda(t),\ldots,\mathbf{r_N}/ \lambda(t) )
\label{eq:ansatz}
\end{equation}
where the time dependent normalisation factor  $\mathcal{N}(t)$ 
and the time dependent
scaling factor $\lambda(t)$ need to be determined.
This ansatz is the combination of a gauge transformation (see the
Gaussian phase factor) and a time dependent 
scaling transform (see the rescaling of
the coordinates by $\lambda(t)$ in $\psi_0$).

The first step is to check that the ansatz Eq.(\ref{eq:ansatz}) is
in the right Hilbert space, i.e.\ that it satisfies the 
boundary conditions Eq.(\ref{eq:clu}).
The scaling transform indeed preserves the boundary conditions: 
it rescales and multiplies the function $A$ by $\lambda$
but does not lead to the appearance of a $O(1)$ term in
Eq.(\ref{eq:clu}). Note that for a non-zero value of $1/a$ the conclusion
would be different if $A\neq 0$.
The gauge transform also preserves the boundary conditions: for
fixed $\mathbf{r_k}$'s we write $r_i^2 + r_j^2  
= 2 R_{ij}^2 +  r_{ij}^2/2$
so that for a fixed $\mathbf{R_{ij} }$, the gauge transform involves
only as a non-constant factor
\begin{equation}
e^{i m r_{ij}^2 \dot\lambda / 4\hbar \lambda }
=1 + i m r_{ij}^2 \dot\lambda / 4\hbar \lambda + O(r_{ij}^4).
\end{equation}
As $r_{ij}^2 \times A/r_{ij} = O(r_{ij})$, the boundary
conditions are preserved by the gauge transform, a conclusion that
extends to the non-zero $1/a$ case.

What is left is to check that the free particle Schr\"odinger equation
Eq.(\ref{eq:overD}) is satisfied by the ansatz for an appropriate
choice of $\lambda(t)$ and $\mathcal{N}(t)$.
One calculates the time derivative and the Laplacian of the ansatz.
One uses the fact that $\psi_0$ is an eigenstate of energy $E$
to express the action of the kinetic energy operator on $\psi_0$
in terms of $r_j^2 \psi_0$ terms and $E \psi_0$.
Equating the terms $r_j^2 \psi_0$ on both sides of Eq.(\ref{eq:overD})
leads to
\begin{equation}
\ddot\lambda(t) =  \frac{\omega_0^2}{\lambda^3(t)} 
-\omega^2(t) \lambda(t)
\end{equation}
 to be solved with the initial conditions $\lambda(t<0)=1$,
$\dot\lambda(t<0)=0$.
For a ballistic expansion, one finds $\lambda(t) = (1+\omega_0^2 t^2)^{1/2}$.
Equating the terms proportional to $\psi_0$ gives
$\dot\mathcal{N}/\mathcal{N}=-3 N\dot \lambda/(2\lambda) 
-i E/(\hbar \lambda^2)$
which is readily integrated in
$\mathcal{N}(t) = \lambda^{-3N/2}(t)
\exp[-i E \int_0^t d\tau/\lambda^{2}(\tau)\hbar]. $
The first factor in the right hand side of this equation
ensures the conservation of the norm of the wavefunction.

The ansatz Eq.(\ref{eq:ansatz})
therefore gives, in the unitary limit, 
the exact time evolution of the
initial state vector for an arbitrary time dependence of the isotropic
harmonic potential.  It rigorously confirms the scaling law that one would
obtain from the zero temperature hydrodynamic approximation.
It also allows to use
the symmetry considerations developed in \cite{Pitaevskii}: e.g.
the limit of linear response to a sudden 
change in the frequency $\omega$ away from
$\omega_0$ gives rise to an undamped oscillation at frequency $2\omega_0$
which reveals the existence of $N$-body stationary states of energy $E\pm 2\hbar\omega_0$
coupled to $\psi_0$ by the excitation operator $\sum_{i=1}^{N} r_i^2$.

We thank L. Pricoupenko for discussions, A. Sinatra, I. Carusotto,
J. Dalibard and C. Cohen-Tannoudji for comments.
LKB is a research unit of \'Ecole normale sup\'erieure and Universit\'e Pierre et Marie,
associated to CNRS.





\begin{thebibliography}{00}
\bibitem{espoir}
It was expected that a gauge plus scaling transform would provide
an exact solution to the $N$-body problem in a strictly two-dimensional
Bose gas in an isotropic harmonic trap \cite{Pitaevskii},
for an interaction potential modeled by
a delta distribution. Actually 
a delta interaction potential in 2D does not lead
to a mathematically well defined problem and the
required regularisation \cite{Olshanii}
breaks the scaling invariance.
\end{thebibliography}
\end{document}